\documentclass[twocolumn,showpacs,preprintnumbers,amsmath,amssymb]{revtex4}
 
\usepackage{graphicx}
\usepackage{dcolumn}
\usepackage{bm}
 
\newcommand{\dpro}[2]{\langle\hspace{-0.06cm}\langle{#1}|{#2}\rangle
\hspace{-0.06cm}\rangle}
\newcommand{\bra}[1]{\langle{#1}|}
\newcommand{\ket}[1]{|{#1}\rangle}
\newcommand{\dbra}[1]{\langle\hspace{-0.06cm}\langle{#1}|}
\newcommand{\dket}[1]{|{#1}\rangle\hspace{-0.06cm}\rangle}

\newcommand{\dgg}{^{\dagger}}
\newcommand{\Tr}{{\rm Tr}\hspace{0.07cm}}

\newcommand{\half}{\frac{1}{2}}

\newcommand{\bmi}{\mbox{\boldmath $I$}}
\newcommand{\bms}{\mbox{\boldmath $S$}}
\newcommand{\bmx}{\mbox{\boldmath $X$}}

\newcommand{\bme}{\mbox{\boldmath $E$}}
\newcommand{\bmr}{\mbox{\boldmath $R$}}

\begin{document}
 
\preprint{APS/123-QED}
 
\title{Suboptimal quantum-error-correcting procedure \\
based on semidefinite programming}
 
\author{Naoki Yamamoto}
 \affiliation{
Control and Dynamical Systems, California
Institute of Technology, Pasadena, California 91125, USA}
 \email{naoki@cds.caltech.edu}
\author{Shinji Hara}%
 \email{Shinji_Hara@ipc.i.u-tokyo.ac.jp}
\author{Koji Tsumura}%
 \email{Koji_Tsumura@ipc.i.u-tokyo.ac.jp}
\affiliation{%
Department of Information Physics and Computing, University of Tokyo,
Hongo 7-3-1, Bunkyo-ku, Tokyo 113-0033, Japan
}%
\date{\today}
 
\begin{abstract}
In this paper, we consider a simplified error-correcting problem: 
for a fixed encoding process, to find a cascade connected quantum channel 
such that the worst fidelity between the input and the output 
becomes maximum. 
With the use of the one-to-one parametrization of quantum channels, 
a procedure finding a suboptimal error-correcting channel based 
on a semidefinite programming is proposed.
The effectiveness of our method is verified by an example 
of the bit-flip channel decoding. 
\end{abstract}
 
\maketitle
 

\section{Introduction}

Quantum error correcting \cite{steane,knill,nielsen,shor} is 
surely a necessary technique to protect quantum states against decoherence 
and unexpected noise in quantum computations 
\cite{shor} or communications \cite{nielsen}. 
An error-correcting procedure is composed of encoding and 
decoding processes; the former is usually done by embedding 
an input in a higher-dimensional Hilbert space, e.g., a single 
qubit is encoded as
\begin{eqnarray}
\label{example-intro}
& & \hspace*{-1em}
     {\mathbb C}^{2}\ni \ket{\phi}=a\ket{0}+b\ket{1}
\nonumber \\ & & \hspace*{0.22em}
        \rightarrow \ket{\phi_{c}}=a\ket{0_{c}}+b\ket{1_{c}}
        :=a\ket{000}+b\ket{111}\in({\mathbb C}^{2})^{\otimes 3}.
\nonumber \\ & & \hspace*{-1em}
\end{eqnarray}
The decoding process denoted by a {\it recovery channel} 
${\cal R}$ is in practice implemented by a combination of 
unitary operations and classical measurements, which is 
generally represented by 
\[
     \ket{\phi_{c}}\bra{\phi_{c}}
      \rightarrow{\cal E}(\ket{\phi_{c}}\bra{\phi_{c}})
      \rightarrow{\cal RE}(\ket{\phi_{c}}\bra{\phi_{c}}).
\]
The superoperator ${\cal E}$ represents the occurrence of errors. 
The procedure of error correcting formulated in 
\cite{knill} is to expand the input space appropriately and to design the 
recovery channel ${\cal R}$ such that the worst fidelity 
between the input and the output becomes maximum, i.e., 
\begin{equation}
\label{problem-intro}
    \min_{\ket{\phi_c}}\bra{\phi_c}
            {\cal R}{\cal E}(\ket{\phi_c}\bra{\phi_c})
                   \ket{\phi_c}\rightarrow{\rm max.}
\end{equation}
Especially, a necessary and sufficient condition for the perfect 
error-correcting, ${\cal RE}(\ket{\phi_{c}}\bra{\phi_{c}})
=\ket{\phi_{c}}\bra{\phi_{c}}$, was given in \cite{knill}. 
When the condition is not fulfilled, however, any analytic ways 
to design the recovery channel have been unknown.

The min-max problem (\ref{problem-intro}) is still hard to obtain 
a global optimal solution for even by means of numerical methods. 
Actually, \cite{reimpell} has replaced the problem 
(\ref{problem-intro}) by the simple maximization problem of 
a special case of {\it Schumacher's entanglement fidelity} 
\cite{schumacher}. 
However, the criterion will be inappropriate for practical 
purposes because it is not based on the worst input.

In spite of the difficulty, this paper proposes a numerical 
method to solve the original min-max problem (\ref{problem-intro}) 
for a fixed encoding process. 
The key idea is the relaxation of the problem to a convex 
optimization one with linear matrix inequality (LMI) constraints, 
i.e., a semidefinite programming (SDP). 
The methodologies of SDP's have been used in many fields including 
{\it control theory} \cite{boyd} and even quantum physics. 
For example, \cite{pablo} successfully applied the method of SDP to 
the test distinguishing entangled from separable quantum states.
We also find \cite{audenaert} made use of an SDP to obtain 
the optimal quantum channel which approximates certain desired 
qubit transformations.

The concrete derivation of the SDP is the following. 
Any quantum channel is one-to-one correspondent to a 
positive semidefinite matrix with a linear equality constraint 
\cite{fujiwara,jami,dariano}, which directly concludes the 
convexity of the set of quantum channels \cite{choi}. 
Hence Eq. (\ref{problem-intro}) can be rewritten as a convex 
optimization problem with respect to the recovery channel. 
Moreover, by relaxing the set of inputs and applying the 
{\it S-procedure} \cite{boyd,yakubovich}, the LMI constraints 
are derived.

The suboptimal recovery channel, which is obtained from the 
SDP derived above, guarantees error correction for the 
worst-case input, unlike \cite{reimpell}. 
Also, for the {\it bit-flip channel}, it will show almost 
the same performance as that of a special error-correcting code, 
the majority-rule code.


\section{Problem formulation}

Let ${\cal H}$ and ${\cal K}$ be $n$-dimensional Hilbert spaces of an 
enlarged input quantum state and the corresponding output, respectively. 
For example, ${\cal H}=({\mathbb C}^2)^{\otimes 3}$ in Eq. 
(\ref{example-intro}). 
In this paper we identify ${\cal K}$ with ${\cal H}$ as they have the same 
dimension. 
We denote ${\cal L}({\cal H})$ the set of all matrices on ${\cal H}$. 
A quantum state $\rho$ belongs to 
${\cal S}({\cal H})=\{\rho\in{\cal L}({\cal H})~|~\rho\dgg=\rho\geq 0, 
~\Tr\rho=1\}$. 
An input-output relation of a quantum state is represented by 
a {\it quantum channel}; an input quantum state $\rho\in{\cal S}({\cal H})$ 
is transformed to 
\[
     \rho'=\sum_{k=1}^{m}E_{k}\rho E_{k}\dgg\in{\cal S}({\cal H}),
\]
where $E_{k}\in{\cal L}({\cal H})$ is a matrix with the size $n\times n$ 
and represents channel properties which are often identical to 
the occurrence of some errors. 
Note that 
\[
     \sum_{k=1}^{m}E_{k}\dgg E_{k}=I
\]
has to hold in order that the output $\rho'$ satisfies 
$\Tr\rho'=1$. 
Let us call the set $\{ E_{k} \}$ the {\it error channel}. 
Note also that an element of the error channel $E_{k}$ and its 
unitary transformation $E'_{k}=\sum_{j}u_{jk}E_{j}$, where 
$U=(u_{jk})$ is a unitary matrix, have the same input-output 
relation as follows: 
\begin{equation}
\label{channel-unitary}
      \sum_{k=1}^{m}E_{k}\rho E_{k}\dgg
            =\sum_{k=1}^{m}E'_{k}\rho E'_{k}\mbox{}\dgg,~
            \forall \rho\in{\cal S}({\cal H}).
\end{equation}
We now explain the problem to be investigated. 
In this paper, for the sake of simplicity we fix the encoding 
process, or equivalently we fix ${\cal H}$, and concentrate on 
the decoding process only. 
An encoded input $\rho=\ket{\phi_{c}}\bra{\phi_{c}}$ is restricted 
into the {\it code space} given by
\[
     {\cal C}:=\Big\{ \ket{\phi_{c}} \in{\cal H}~\Big|~
        \ket{\phi_{c}}=\sum_{k=1}^{L}\lambda_{k}\ket{k_{c}} \Big\}
           \subset{\cal H},
\]
where $\{ \ket{k_{c}} \}$ is an orthonormal system in ${\cal H}$. 
In example (\ref{example-intro}), $\ket{1_c}=\ket{000}$ and 
$\ket{2_c}=\ket{111}$ are chosen. 
The decoding process is represented by the {\it recovery channel} 
$\{R_{k}\}$ satisfying $\sum_{k=1}^{M}R_{k}\dgg R_{k}=I$, where 
the number of the elements $R_{k}$, $M$, needs not to be equal 
to that of the error channel $\{E_{k}\}$. 
The recovery channel $\{R_{k}\}$ is just connected to the error 
channel $\{E_{k}\}$, and it is designed so that the output given 
by 
\begin{equation}
\label{channel-withR}
     \rho'=\sum_{k=1}^{M}\sum_{j=1}^{m}R_{k}E_{j}\rho E_{j}\dgg R_{k}\dgg
\end{equation}
is as close to the input $\rho=\ket{\phi_{c}}\bra{\phi_{c}}$ as possible. 
It is notable that the optimal set of matrices $\{R_{k}\}$ is 
not unique even if they exist, because $\{ R_{k} \}$ has 
unitary freedom as seen in Eq. (\ref{channel-unitary}). 
The error-correcting problem, which was originally considered in 
\cite{knill}, is addressed by 
\begin{eqnarray}
& & \hspace*{-1em}
\label{ec-problem}
     \max_{\{ R_{k} \}}~F(\{R_{k}\})
\nonumber \\ & & \hspace*{-1em}
     F(\{R_{k}\}):=\min_{\ket{\phi_{c}}\in{\cal C}}
                                  \bra{\phi_{c}}\rho'\ket{\phi_{c}},
\end{eqnarray}
where $\rho'$ is given by Eq. (\ref{channel-withR}). 
The difference between the input $\rho=\ket{\phi_{c}}\bra{\phi_{c}}$ 
and the output $\rho'$ is quantified by the fidelity 
$f:=\bra{\phi_{c}}\rho'\ket{\phi_{c}}$, which is bounded 
by $0\leq f\leq 1$. 
It is known that the fidelity becomes $1$ if and only if 
$\rho'=\ket{\phi_{c}}\bra{\phi_{c}}$. 
The minimization with respect to the input $\ket{\phi_{c}}$ means 
that the performance of the recovery channel is evaluated in the worst case. 
The best performance $F=1$ is attained if and only if the following 
condition holds.

{\it Theorem 1} \cite{knill}. 
There exists a set $\{R_{k}\}$ satisfying $F(\{R_{k}\})=1$, 
if and only if 
\begin{equation}
\label{theorem1}
      \bra{i_{c}}E_{j}\dgg E_{k}\ket{\ell_{c}}=\alpha_{jk} \delta_{i\ell}
\end{equation}
holds for all $i, j, k, \ell$.

In Eq. (\ref{theorem1}), $\alpha_{jk}$ is a constant to be determined by 
choices of $E_{j}$ and $E_{k}$, and $\delta_{i\ell}$ is Kronecker's delta. 
A remarkable feature of the theorem is that condition (\ref{theorem1}) 
is written by using the properties of only the error channel $\{E_{k}\}$ 
and the code space ${\cal C}$. 
However, Theorem 1 does not tell us any way to search an optimal or even a 
good recovery channel when the condition (\ref{theorem1}) is not satisfied. 
The main contribution of this paper is to present a procedure 
to overcome this critical drawback.


\section{One-to-one parametrization of quantum channels}

It is known that any quantum channel from ${\cal S}({\cal H})$ to 
${\cal S}({\cal H})$ is in one-to-one correspondence with 
a positive semidefinite matrix acting on ${\cal H}^{\otimes 2}$. 
We first restate this fact by a slightly different way from the 
conventional one in \cite{dariano}.

Fixing an orthonormal basis $\{\ket{i}\otimes\ket{j}\}$, 
we denote a vector in ${\cal H}^{\otimes 2}$ by 
$\dket{\Phi}=\sum_{i,j=1}^{n}\lambda_{ij}\ket{i}\otimes\ket{j}$. 
We sometimes write $\ket{i}\ket{j}$ instead of 
$\ket{i}\otimes\ket{j}$ for simplicity. 
Let us introduce a specific vector in ${\cal H}^{\otimes 2}$ given by
\begin{equation}
\label{1vector}
    \dket{{\rm e}}:=\sum_{k=1}^{n}\ket{k}\otimes\ket{k}^{*},
\end{equation}
where the asterisk denotes complex conjugation of each 
element \cite{arimitsu}. 
Then $\dket{{\rm e}}$ is independent of the selection of 
orthonormal basis; i.e., 
\begin{equation}
\label{e-invariant} 
    \dket{{\rm e}}
    =\sum_{k}\ket{a_{k}}\otimes\ket{a_k}^{*}
    =\sum_{k}\ket{b_{k}}\otimes\ket{b_k}^{*}
\end{equation}
holds for any orthonormal basis $\{\ket{a_{k}}\}$ and 
$\{\ket{b_{k}}\}$. 
We should remark that the above fine property is not satisfied 
if $\dket{{\rm e}}$ is defined via the conventional way 
\cite{dariano} as 
$\dket{{\rm e}}=\sum_{k=1}^{n}\ket{k}\otimes\ket{k}$. 
The vector $\dket{{\rm e}}$ also has an important property 
expressed as 
\begin{equation}
\label{TFDformula}
     (A \otimes I)\dket{{\rm e}}=(I \otimes A^{\mathsf T})\dket{{\rm e}},
     ~~\forall A\in{\cal L}({\cal H}),
\end{equation}
where ${\mathsf T}$ denotes the matrix transpose. 
Actually, expanding a matrix $A\in{\cal L}({\cal H})$ by using an 
orthonormal basis $\{ \ket{k} \}$ as $A=\sum_{i,j}a_{ij}\ket{i}\bra{j}$, 
the left hand-side of Eq. (\ref{TFDformula}) becomes 
\[
      \Bigr[\sum_{i,j}a_{ij}\ket{i}\bra{j}\otimes I\Bigr]
      \sum_{k}\ket{k}\ket{k}^{*}
    =\sum_{i,j}a_{ij}\ket{i}\ket{j}^{*}.
\]
Similarly, the right-hand side of Eq. (\ref{TFDformula}) is calculated as 
\[
      \Bigr[I\otimes\sum_{i,j}a_{ij}(\ket{j}\bra{i})^{*}\Bigr]
      \sum_{k}\ket{k}\ket{k}^{*}
    =\sum_{i,j}a_{ij}\ket{i}\ket{j}^{*},
\]
which implies the equality (\ref{TFDformula}).

Here we define a positive semidefinite matrix 
$\bmx_1\in{\cal L}({\cal H}^{\otimes 2})$ associated with a quantum channel 
$\rho'=\sum_k X_k\rho X_k\dgg$ as 
\[
    \bmx_1:=\sum_k(X_k\otimes I)\dket{{\rm e}}\dbra{{\rm e}}
                                (X_k\dgg\otimes I). 
\]
It turns out that the trace-preserving condition $\sum_k X_k\dgg X_k=I$ 
corresponds to $(\Tr_{\cal H}\otimes{\rm id})\bmx_1=I$, and the map 
$\rho\rightarrow\rho'$ is written in terms of $\bmx_1$ as 
\begin{equation}
\label{channel-2}
    \rho'=({\rm id}\otimes\Tr_{\cal H})
               \Big[(I\otimes\rho^{\mathsf T})\bmx_1\Big].
\end{equation}
Here, ${\rm id}$ denotes the identity operator on ${\cal H}$. 
Conversely, it is known that any quantum channel can be 
represented by using a positive semidefinite matrix as the 
above form (\ref{channel-2}) \cite{dariano}.
That is, Eq. (\ref{channel-2}) defines a one-to-one correspondence 
between a quantum channel and a positive semidefinite matrix on 
${\cal H}^{\otimes 2}$. 
Actually, $\bmx_1$ is invariant under the unitary transformation 
of the structuring matrices $X'_{k}=\sum_{j}u_{jk}X_{j}$, 
where $U=(u_{jk})$ is a unitary matrix, i.e., 
\begin{eqnarray}
& & \hspace*{-1em}
\label{unitary-invariant-vec}
    \bmx_1
     =\sum_k(X_k\otimes I)\dket{{\rm e}}\dbra{{\rm e}}(X_k\dgg\otimes I)
\nonumber \\ & & \hspace*{0.8em}
     =\sum_k(X'_k\otimes I)\dket{{\rm e}}
                           \dbra{{\rm e}}(X'_k\mbox{}\dgg\otimes I).
\end{eqnarray}

We next introduce another matrix expression for a quantum 
channel, $\bmx_2\in{\cal L}({\cal H}^{\otimes 2})$. 
The introduction of $\bmx_2$ in addition to $\bmx_1$ is 
essential from the computational viewpoint, which will be 
explained in the last part of Section V. 
Let us define a vector associated with a quantum state 
$\rho\in{\cal S}({\cal H})$ as 
\begin{equation}
\label{def-of-vecrho}
    \dket{\rho}:=(\rho \otimes I)\dket{{\rm e}}
                   \in{\cal H}^{\otimes 2},
\end{equation}
which is obviously in one-to-one correspondence with $\rho$. 
In particular, the vector representation of a pure state 
$\rho=\ket{\phi}\bra{\phi}$ is given by 
\begin{equation}
\label{pure}
    \dket{\rho}=(\ket{\phi}\bra{\phi}\otimes I)
                    \sum_{k}\ket{k}\otimes\ket{k}^{*}
               =\ket{\phi}\otimes\ket{\phi}^{*},
\end{equation}
since $\dket{{\rm e}}$ is independent of the selection of $\{\ket{k}\}$. 
Multiplying an input-output relation of a quantum channel, 
$\rho'=\sum_{k}X_{k}\rho X_{k}\dgg$, by $\dket{{\rm e}}$ from the right, 
we have $(\rho'\otimes I)\dket{{\rm e}}
=\sum_{k}(X_{k}\rho X_{k}\dgg\otimes I)\dket{{\rm e}}$. 
On account of Eq. (\ref{TFDformula}), this relation becomes 
\[
     (\rho'\otimes I)\dket{{\rm e}}
       =\sum_{k}(X_{k}\otimes X^{*}_{k})(\rho\otimes I)\dket{{\rm e}}.
\]
From the definition (\ref{def-of-vecrho}), this equation 
is described by 
\[
      \dket{\rho'}=\bmx_2\dket{\rho},
\]
where $\dket{\rho'}=(\rho'\otimes I)\dket{{\rm e}}$ and 
\[
     \bmx_2:=\sum_{k}X_{k}\otimes X^{*}_{k}.
\]
The trace-preserving condition is rewritten by 
\[
    \dbra{{\rm e}}\bmx_2=\dbra{{\rm e}}\sum_{k}X_k\otimes X^{*}_k
                 =\dbra{{\rm e}}\sum_{k}I\otimes X^{{\mathsf T}}_k X^{*}_k
                 =\dbra{{\rm e}}.
\]
Similar to $\bmx_1$, we see that $\bmx_2$ is also invariant 
under the unitary transformation of the structuring matrices 
$X'_{k}=\sum_{j}u_{jk}X_{j}$. 
The matrix $\bmx_2$ is associated with $\bmx_1$ through the 
rearrangement of the elements as 
\[
     \bra{i}\bra{j}^{*}\bmx_2\ket{k}\ket{\ell}^{*}
      =\bra{i}\bra{k}^{*}\bmx_1\ket{j}\ket{\ell}^{*}.
\]
This relation is independent of the selection of $\{\ket{k}\}$ 
because of the property (\ref{e-invariant}). 
Since the rearrangement map is obviously homeomorphism, 
the set of $\bmx_1$ is equivalent to that of $\bmx_2$. 
Denoting this relation by $\bmx_1=\Phi(\bmx_2)$, 
the above discussions are summarized as follows.

{\it Theorem 2.}
Any input-output relation of a quantum state is written by using Eq. 
(\ref{def-of-vecrho}) as $\dket{\rho'}=\bmx\dket{\rho}$, 
where the matrix $\bmx$ is included in the set 
\[
     {\cal X}=\{~\bmx\in{\cal L}({\cal H}^{\otimes 2}) ~|~
      \Phi(\bmx)\geq 0,~\dbra{{\rm e}}\bmx=\dbra{{\rm e}}~\}. 
\]
The transformation $\Phi(\bmx)$ is defined with respect to 
an orthonormal basis $\{\ket{k}\}$ as 
\[
     \bra{i}\bra{j}^{*}\bmx\ket{k}\ket{\ell}^{*}
      =\bra{i}\bra{k}^{*}\Phi(\bmx)\ket{j}\ket{\ell}^{*}.
\]
The set ${\cal X}$ is convex, and its dimension is $n^{4}-n^{2}=N^{2}-N$. 

The convexity of ${\cal X}$ is obvious. 
We also note that for $\bme_{1}\in{\cal X}$ and 
$\bme_{2}\in{\cal X}$, we have $\bme_{1}\bme_{2}\in{\cal X}$ and 
$\bme_{2}\bme_{1}\in{\cal X}$, which reflect that the fact the cascade 
connection of quantum channels is also a quantum channel.


\section{Recovery channel design based on semidefinite programming}

Let us rewrite the problem (\ref{ec-problem}) by using the vector 
representation of a quantum state (\ref{def-of-vecrho}). 
As seen in Eq. (\ref{pure}), the pure input 
$\rho=\ket{\phi_{c}}\bra{\phi_{c}}$ is represented by 
$\dket{\rho}=\ket{\phi_{c}}\ket{\phi_{c}}^{*}$. 
The output through the error channel $\bme$ and the recovery 
channel $\bmr$ is given by $\dket{\rho'}=\bmr\bme\dket{\rho}=
\bmr\bme\ket{\phi_{c}}\ket{\phi_{c}}^{*}$.
Then, the inner product of $\dket{\rho}$ and $\dket{\rho'}$ 
is calculated as 
\begin{eqnarray}
& & \hspace*{-1em}
      \dpro{\rho}{\rho'}=
      \bra{\phi_{c}}\bra{\phi_{c}}^{*}
              \bmr\bme\ket{\phi_{c}}\ket{\phi_{c}}^{*}
\nonumber \\ & & \hspace*{2.1em}
      =\sum_{j,k}|\bra{\phi_{c}}R_{j}E_{k}\ket{\phi_{c}}|^{2}
      =\bra{\phi_{c}}\rho'\ket{\phi_{c}},
\nonumber
\end{eqnarray}
which is just the fidelity between the input and output. 
In view of this, the problem (\ref{ec-problem}) is rewritten as 
\begin{eqnarray}
& & \hspace*{-1em}
\label{ec-problem-vec}
     \max_{R\in{\cal X}}~F(\bmr)
\nonumber \\ & & \hspace*{-1em}
     F(\bmr):=\min_{\ket{\phi_{c}}\in{\cal C}}
                   \bra{\phi_{c}}\bra{\phi_{c}}^{*}
                        \bmr\bme\ket{\phi_{c}}\ket{\phi_{c}}^{*}. 
\end{eqnarray}
The optimal recovery channel $\bmr_{o}$ of the above problem is uniquely 
determined, while in the original problem 
(\ref{ec-problem}), the optimal set of the matrices $\{R_{k}\}$ 
cannot be determined uniquely due to the unitary freedom 
(\ref{channel-unitary}).

Before describing the procedure to find a suboptimal recovery 
channel, we derive a necessary condition for the perfect recovery channel 
to exist. 
The $(i\otimes i, j\otimes j)$ element of $\bme\dgg\bme$ is calculated as 
\begin{eqnarray}
& & \hspace*{-1em}
     \bra{i_{c}}\bra{i_{c}}^{*}
              \bme\dgg\bme\ket{j_{c}}\ket{j_{c}}^{*}
\nonumber \\ & & \hspace*{0em}
     =\bra{i_{c}}\bra{i_{c}}^{*}
       \Big[ \sum_{k,\ell}(E\dgg_{k}E_{\ell})\otimes
                          (E\dgg_{k}E_{\ell})^{*}\Big]
      \ket{j_{c}}\ket{j_{c}}^{*}
\nonumber \\ & & \hspace*{0em}
     =\sum_{k,\ell}|\bra{i_{c}}E\dgg_{k}E_{\ell}\ket{j_{c}}|^{2}.
\nonumber
\end{eqnarray}
When the perfect recovery channel exists, the relation 
$\bra{i_{c}}E\dgg_{k}E_{\ell}\ket{j_{c}}
=\alpha_{k\ell}\delta_{ij}$ has to hold for all $i, j, k, \ell$ 
from Theorem 1, which leads to 
\begin{equation}
\label{perfect-check}
     \bra{i_{c}}\bra{i_{c}}^{*}
              \bme\dgg\bme\ket{j_{c}}\ket{j_{c}}^{*}
     =\sum_{k,\ell}|\alpha_{k\ell}\delta_{ij}|^{2}:=\alpha\delta_{ij}.
\end{equation}
Hence a perfect recovery channel never exists under the condition 
$\bra{i_{c}}\bra{i_{c}}^{*}\bme\dgg\bme\ket{j_{c}}\ket{j_{c}}^{*}
\neq\alpha\delta_{ij}$, which is easy to check.

Now we shall give the procedure to find a suboptimal recovery channel 
when the condition (\ref{perfect-check}) does not hold. 
First note that the original problem (\ref{ec-problem-vec}) is equivalent 
to the following minimization problem: 
\[
     \min_{R\in{\cal X}}~\epsilon~~\mbox{s.t.}~~
     \bra{\phi_{c}}\bra{\phi_{c}}^{*}
              \bmr\bme\ket{\phi_{c}}\ket{\phi_{c}}^{*}
              > 1-\epsilon,~
              \forall \ket{\phi_{c}}\in{\cal C}.
\]
We then relax the problem into 
\begin{equation}
\label{relaxed-final}
     \min_{R\in{\cal X}}~\epsilon~~\mbox{s.t.}~~
              \dbra{\phi}\bmr\bme\dket{\phi} > 1-\epsilon,~
              \forall \dket{\phi}\in{\cal C}\otimes{\cal C}^*, 
\end{equation}
where ${\cal C}^*$ is defined as a linear subspace spanned by 
$\{\ket{k_c}^*\}$. 
The meaning of the relaxation is the following: 
any element in
${\cal C}\otimes{\cal C}^*\subset{\cal H}^{\otimes 2}$ 
is always given by 
$\dket{\phi}=\sum_{j,k=1}^{L}\phi_{jk}\ket{j_c}\ket{k_c}^*$, 
where the coefficient $\phi_{jk}\in{\mathbb C}$ has no restriction. 
Therefore, the original input $\ket{\phi_{c}}\ket{\phi_c}^* 
=\sum_{j,k=1}^{L}\lambda_{j}\lambda_{k}^*\ket{j_c}\ket{k_c}^*$ 
is obviously included in ${\cal C}\otimes{\cal C}^*$, which 
indicates that the input is allowed to be a linear operator of 
the form $\phi=\sum_{j,k=1}^{L}\phi_{jk}\ket{j_{c}}\bra{k_{c}}$ 
in addition to the pure state $\rho=\ket{\phi_{c}}\bra{\phi_{c}}$ 
in the relaxed problem (\ref{relaxed-final}). 
The relaxed condition is equivalent to the following inequality: 
\begin{eqnarray}
& & \hspace*{-1em}
\label{condition-pre2}
     \dbra{\phi}\Big[ \half(\bmr\bme)+\half(\bmr\bme)\dgg
             +(\epsilon-1)\bmi\Big]\dket{\phi}>0,
\nonumber \\ & & \hspace*{12em}
     \forall \dket{\phi}\in{\cal C}\otimes{\cal C}^*. 
\end{eqnarray}

Now we utilize a famous formula named the {\it S-procedure} 
\cite{boyd,yakubovich};
\begin{eqnarray}
& & \hspace*{-1em}
     \dbra{x}\bmx\dket{x}>0, \hspace*{0.5em}
     \forall \dket{x} \in {\cal S}:=\{ \dket{x}\neq 0 ~|~
     \dbra{x}\bms\dket{x}\geq 0 \}
\nonumber \\ & & \hspace*{-1em}
     ~\Leftrightarrow~
     \exists \tau>0~~\mbox{s.t.}~~\bmx-\tau \bms>0.
\nonumber
\end{eqnarray}
Note that $\bmx$ and $\bms$ need not be positive. 
This formula says that we may find at least one positive 
number $\tau$ such that $\bmx-\tau \bms>0$ is satisfied 
instead of checking $\dbra{x}\bmx\dket{x}>0$ for all 
$\dket{x}\in{\cal S}$.

We apply the S-procedure to obtain an equivalent relation to 
the relaxed condition (\ref{condition-pre2}). 
Assume $\bms$ is a negative semidefinite matrix acting on 
$({\cal C}\otimes{\cal C}^*)^{\perp}$. 
Then, all vectors $\dket{\phi}\in{\cal C}\otimes{\cal C}^*$ 
satisfy $\dbra{\phi}\bms\dket{\phi}=0$, and the condition 
(\ref{condition-pre2}) is equivalently transformed to 
\[
      \exists \tau>0,~~\mbox{s.t.}~~
      \half(\bmr\bme)+\half(\bmr\bme)\dgg
             +(\epsilon-1)\bmi-\tau \bms>0.
\]
Consequently, our relaxed problem is to minimize $\epsilon$ 
subject to the following LMIs: 
\begin{eqnarray}
& & \hspace*{-1em}
\label{LMI-1}
     \half(\bmr\bme)+\half(\bmr\bme)\dgg
             +(\epsilon-1)\bmi-\tau \bms>0,
\\ & & \hspace*{-1em}
\label{LMI-2}
     \Phi(\bmr)\geq 0, ~\tau>0,
\\ & & \hspace*{-1em}
\label{LMI-3}
     \dbra{{\rm e}}\bmr=\dbra{{\rm e}},
\end{eqnarray}
with changing $\bms\in{\cal L}(({\cal C}\otimes{\cal C}^*)^{\perp})$. 
This is a typical SDP, which enables us to find a suboptimal solution 
$\bmr_{o}$ achieving the maximum fidelity $F(\bmr_{o})$.


\section{Application to the bit flip channel decoding}

Let us consider the bit-flip channel with flipping 
probability $p$: 
\[
    T(\rho)=p\sigma_{x}\rho\sigma_{x}+q\rho,
\]
where $p+q=1$ and $\sigma_{x}=\ket{0}\bra{1}+\ket{1}\bra{0}$. 
When any encoding and decoding processes are not performed, 
the minimum fidelity between a pure input state and the output is 
$q$. 
In this section, we consider two kinds of encoded input: 
one is $\ket{\phi_c}=a\ket{1_c}+b\ket{2_c}:=a\ket{00}+b\ket{11}$, 
and the other is 
$\ket{\phi_c}=a\ket{1_c}+b\ket{2_c}:=a\ket{000}+b\ket{111}$. 
They are perturbed by the error channels $T^{\otimes 2}$ and 
$T^{\otimes 3}$, respectively. 
For the first case, we will examine the gap between the 
overguaranteed fidelity due to the relaxation and original 
minimum fidelity. 
For the second, we compare our numerical procedure with a famous 
error-correcting strategy, the {\it majority-rule code}.

For the error channel $T^{\otimes 2}$, the interaction operators are 
given by 
\begin{eqnarray}
& & \hspace*{-2em}
    E_1=p\sigma_x\otimes\sigma_x,~~E_2=\sqrt{pq}\sigma_x\otimes I,
\nonumber \\ & & \hspace*{-2em}
    E_3=\sqrt{pq}I \otimes\sigma_x,~~E_4=qI\otimes I.
\nonumber
\end{eqnarray}
The corresponding matrix $\bme=\sum_{k=1}^{4}E_{k}\otimes E^{*}_{k}
\in{\cal L}({\cal H}^{\otimes 2})$ is expressed as 
\[
   \bme=\left[ \begin{array}{cc|cc}
         qE_4 & \sqrt{pq}E_3 & \sqrt{pq}E_2 & pE_1 \\
         \sqrt{pq}E_3 & qE_4 & pE_1 & \sqrt{pq}E_2 \\ \hline
         \sqrt{pq}E_2 & pE_1 & qE_4 & \sqrt{pq}E_3 \\
         pE_1 & \sqrt{pq}E_2 & \sqrt{pq}E_3 & qE_4
        \end{array} \right].
\]
We first get
\[
    \bra{1_{c}}\bra{1_{c}}^{*}\bme\dgg\bme\ket{2_{c}}\ket{2_{c}}^{*}
     =3p^2q^2+p^4\neq 0,
\]
which violates the condition (\ref{perfect-check}); thus, there does 
not exist a perfect recovery channel. 
Now ${\cal C}\otimes{\cal C}^{*}$ is spanned by 
$\{\ket{1_{c}}\ket{1_{c}}^{*}, \ket{1_{c}}\ket{2_{c}}^{*}, 
   \ket{2_{c}}\ket{1_{c}}^{*}, \ket{2_{c}}\ket{2_{c}}^{*} \}$, 
and hence we can take 
$\bms\in{\cal L}(({\cal C}\otimes{\cal C}^*)^{\perp})$ 
as the following negative semidefinite matrix: 
\begin{equation}
       \bms=\left[ \begin{array}{cc|cc}
             -I' & & & \\
              & -I_{4} & & \\ \hline
              & & -I_{4} & \\
              & & & -I'
         \end{array} \right],
\nonumber
\end{equation}
where $I_{4}$ denotes the $4\times 4$ identity matrix and 
$I':={\rm diag}\{ 0, 1, 1, 0 \}$.

Consider the case $p=9/10$. 
By using MATLAB LMI Toolbox, the set of LMI's (\ref{LMI-1}), 
(\ref{LMI-2}), and (\ref{LMI-3}) is solved for the minimum 
error $\epsilon=0.196$ with a suboptimal solution 
\begin{eqnarray}
& & \hspace*{-1em}
\label{suboptimal}
   \bmr_{o}=
\nonumber \\ & & \hspace*{-1em}
   \left[ \begin{array}{cc|cc}
     O&\alpha B_{12}+\beta B_{43}&\alpha B_{13}+\beta B_{42}&B_{14}+B_{41} \\
     O & \gamma B_{22}         & \gamma B_{23}         & O \\ \hline
     O & \gamma B_{32}         & \gamma B_{33}         & O \\
     B_{14}+B_{41}&\alpha B_{42}+\beta B_{13}&\alpha B_{43}+\beta B_{12}&O
   \end{array} \right],
\nonumber \\ & & \hspace*{-2em}
\end{eqnarray}
where $B_{ij}$ is a $4\times 4$ matrix whose $(i,j)$ element 
is $1$ and others zero. 
The parameters are given by $\alpha=0.28, \beta=0.09$ and 
$\gamma=0.22$, and $O$ denotes the $4\times 4$ zero matrix. 
We here round all entries of $\bmr_{o}$ off to two decimal 
places. 
Note that Eq. (\ref{suboptimal}) is one of the suboptimal recovery channels 
derived from the relaxed problem (\ref{relaxed-final}). 
That is, Eq. (\ref{suboptimal}) guarantees the minimum error 
$\epsilon=0.196$ for extra inputs which are taken outside 
the code space.

Let us compute $F(\bmr_{o})$, the actual fidelity for the above 
suboptimal recovery channel $\bmr_{o}$, in order to check the 
gap between the optimal one and the suboptimal one obtained 
by the proposed method. 
An arbitrary input in ${\cal C}$ is described by 
\[
         \ket{\phi_{c}}=a\ket{1_{c}}+b\ket{2_{c}}=
                     \left[ \begin{array}{c}
                        a \\
                        0 \\
                        0 \\
                        b \\
                     \end{array} \right], \hspace*{0.4em}
                    |a|^{2}+|b|^{2}=1.
\]
For the error channel without any recovery, the fidelity between 
the input and output is 
\begin{eqnarray}
& & \hspace*{-1em}
    \bra{\phi_{c}}\bra{\phi_c}^*\bme
         \ket{\phi_{c}}\ket{\phi_c}^*
\nonumber \\ & & \hspace*{0em}
    =q^2+2p^2\Big[|a^{*}b|^{2}+\Re(a^{*2}b^{2})\Big]\geq 0.01,
\nonumber
\end{eqnarray}
where $\Re$ denotes the real part. 
This implies that the minimum fidelity is $0.01$. 
Also, the minimum fidelity without any encoding and 
decoding processes is $q=0.1$. 
Thus we can assert that our recovery channel (\ref{suboptimal}) 
works quite well because it certifies the fidelity 
$1-\epsilon=1-0.196=0.804$ at least. 
Now the input-output fidelity for the error channel accompanied 
with the recovery channel (\ref{suboptimal}) is calculated by 
\begin{eqnarray}
& & \hspace*{-2em}
      \bra{\phi_{c}}\bra{\phi_c}^*\bmr_o\bme
         \ket{\phi_{c}}\ket{\phi_c}^*
\nonumber \\ & & \hspace*{-1em}
          =0.56pq+p^2
           +2(q^2+0.18pq)\Big[|a^{*}b|^{2}+\Re(a^{*2}b^{2})\Big]
\nonumber \\ & & \hspace*{-1em}
      \geq 0.56pq+p^2 \cong 0.860,
\nonumber
\end{eqnarray}
where the equality holds when $a=1$ or $b=1$. 
This actual minimum value is slightly larger than the guaranteed 
worst value $1-\epsilon=0.804$. 
This difference appears for the following reason: the input 
$\dket{\phi}\in{\cal C}\otimes{\cal C}^*$ corresponding to the worst 
case is selected outside the code space ${\cal C}$ for the relaxed 
problem (\ref{relaxed-final}), while the actual worst input is shown 
to be $\ket{\phi_{c}}=\ket{1_{c}}$ or $\ket{\phi_{c}}=\ket{2_{c}}$ 
in ${\cal C}$. 
However, we may conclude that the recovery channel (\ref{suboptimal}) 
is not so conservative because the difference is not too large.

The next subject is to compare our numerical procedure with 
the (improved) majority rule code, e.g., 
$\ket{001}\rightarrow\ket{0},~\ket{011}\rightarrow\ket{1}$ 
for $0<p\leq 1/2$ and 
$\ket{001}\rightarrow\ket{1},~\ket{011}\rightarrow\ket{0}$ 
for $1/2<p<1$. 
An input state $\ket{\phi}=a\ket{0}+b\ket{1}$ is encoded into 
$\ket{\phi_c}=a\ket{000}+b\ket{111}$ and passes through the 
triple bit-flip channel $T^{\otimes 3}$. 
When $0<p\leq 1/2$, the output is decoded into 
\[
    \rho'=p^2(3-2p)\sigma_x\ket{\phi}\bra{\phi}\sigma_x
          +q^2(1+2p)\ket{\phi}\bra{\phi}. 
\]
It turns out that the fidelity satisfies 
\begin{eqnarray}
& & \hspace*{-2em}
    F_{{\rm maj}}:=\bra{\phi}\rho'\ket{\phi}
\nonumber \\ & & \hspace*{-1em}
     =q^2(1+2p)+2p^2(3-2p)
                     \Big[|a|^2|b|^2+\Re(a^{*2}b^2)\Big]
\nonumber \\ & & \hspace*{-1em}
                  \geq q^2(1+2p), 
\nonumber
\end{eqnarray}
where the equality is attained when $a=1$ or $b=1$. 
Similarly, we have $F_{{\rm maj}}\geq p^2(3-2p)$ when $1/2<p<1$. 
On the other hand, we shall solve the LMI's (\ref{LMI-1}), 
(\ref{LMI-2}), and (\ref{LMI-3}) with the negative 
semidefinite matrix $\bms={\rm diag}\{0,-1,-1,\ldots\}$, 
where zeros appear in the $1{\rm st}, 8{\rm th}, 57{\rm th}$, 
and $64{\rm th}$ entries, and the others are all $-1$. 
Let us consider the case $p=9/10$ again. 
Then the minimum fidelity via the majority-rule code is given by 
${\rm min}F_{{\rm maj}}=0.972$, whereas our numerical method 
yields a suboptimal recovery channel with the minimum error 
$\epsilon=0.048$, or equivalently, the worst fidelity 
$1-\epsilon=0.952$, i.e., 
\[
    \dbra{\phi}\bmr_o\bme\dket{\phi}\geq 0.952,~~
       \forall \dket{\phi}\in{\cal C}\otimes{\cal C}^*. 
\]
The guaranteed fidelity is slightly less than ${\rm min}F_{{\rm maj}}$ 
due to the relaxation. 
For the other case $p=1/10$, analogously, we have 
${\rm min}F_{{\rm maj}}=0.972$ while the LMI's are solved for the 
same worst fidelity $1-\epsilon=0.952$. 
The above investigations conclude that the proposed method, which 
is applicable for general quantum channel without any prior 
knowledge, has almost the same performance as that of a special 
code for the bit-flip channel.

We lastly remark on an important reason why $\bmr$ is 
introduced in addition to $\Phi(\bmr)$. 
Actually, for the problem in \cite{audenaert}, i.e., 
approximating certain desired qubit transformations via a 
quantum channel, the solution does not depend on whether 
we use $\bmr$ to describe the LMI's or not. 
Now the fidelity to be optimized can be described 
in terms of $\Phi(\bmr)$ as 
\[
    F=\min_{\ket{\phi_c}\in{\cal C}}
                  \bra{\phi_c}\bra{\phi_c}^*
                    \sum_k(I\otimes E_k^{{\mathsf T}})\Phi(\bmr)
                       (I\otimes E_k^*)\ket{\phi_c}\ket{\phi_c}^*, 
\]
and thus the same procedure as in Section IV leads to an LMI: 
\begin{equation}
\label{LMI-PR}
     \sum_k(I\otimes E_k^{{\mathsf T}})\Phi(\bmr)
                       (I\otimes E_k^*)
           +(\epsilon-1)\bmi-\tau\bms>0. 
\end{equation}
However, the solution $\bmr'_o$ obtained from Eqs. (\ref{LMI-2}) and 
(\ref{LMI-PR}) and $(\Tr_{{\cal H}}\otimes{\rm id})\Phi(\bmr)=I$ is 
no longer the same as the one via the LMI's (\ref{LMI-1}), 
(\ref{LMI-2}), and (\ref{LMI-3}), because the relaxed constraints 
disagree each other, i.e., 
\[
    \dbra{\phi}\sum_k(I\otimes E_k^{{\mathsf T}})\Phi(\bmr)
                       (I\otimes E_k^*)\dket{\phi}
    \neq \dbra{\phi}\bmr\bme\dket{\phi}. 
\]
Indeed, for the double bit flip channel $T^{\otimes 2}$ with 
$p=9/10$, the LMI's including Eq. (\ref{LMI-PR}) provide a solution for 
the minimum error $\epsilon=0.749$, whereas $\epsilon=0.196$ via 
the LMI's (\ref{LMI-1}), (\ref{LMI-2}), and (\ref{LMI-3}). 
For this reason, unlike \cite{audenaert}, the introduction of 
$\bmr$ in addition to $\Phi(\bmr)$ is essential in our problem 
formulation.


\section{Conclusion}

In this paper, we have considered a simplified error-correcting problem: 
for a fixed encoding process, to find a cascade-connected quantum channel 
such that the worst fidelity between the input and output becomes maximum. 
With the use of the one-to-one parametrization of quantum channel 
\cite{fujiwara,jami,dariano}, a suboptimal recovery channel can be 
determined as a solution of a semidefinite programming. 
The effectiveness of the proposed method has been verified by 
studying the bit-flip channel. 
Although we could find a suboptimal recovery channel which is 
very close to the optimal one for the example, the condition 
where a good recovery channel exists has not been cleared yet, 
which is an important future work.


\mbox{}

\begin{acknowledgements}

We would like to thank T. Ogawa for his useful comments. 
This work was supported in part by JSPS Grants-in-Aid No.0310897. 

\end{acknowledgements}

\end{document}